\documentclass[pre,superscriptaddress,aps, bibliography, twocolumn, reprint, showpacs, footnoteinbib]{revtex4-1}

\usepackage{graphicx}
\usepackage{amssymb}   
\usepackage{amsfonts}
\usepackage{amsmath}
\usepackage{dsfont}
\usepackage{dcolumn}   
\usepackage{bm}        

\usepackage[utf8]{inputenc}
\usepackage{tikz} 
\usepackage{graphicx}
\usepackage{amssymb}   
\usepackage{amsfonts}
\usepackage{amsmath}
\usepackage{dsfont}
\usepackage{natbib}
\usepackage{bm}        

\usepackage{mathtools}
\usepackage{dsfont}
\usepackage{hyperref}
\usepackage{setspace}

\hypersetup{
    colorlinks = true,
    linkcolor  = blue,
    citecolor  = blue,
    urlcolor   = blue,
}
\definecolor{myblue}{rgb}{0,0,0.75}

\usepackage{graphicx}
\usepackage{amsmath,amssymb,bm}
\usepackage{enumerate}
\usepackage{braket}        
\usepackage{physics}
\usepackage{microtype}

\graphicspath{ {./img/} }

\renewcommand{\vec}[1]{{\boldsymbol{#1}}}
\DeclareMathOperator{\IPR}{IPR}

\begin{document}

\title{Effects of critical correlations on quantum percolation in two dimensions}

\author{Giuseppe De Tomasi}
\affiliation{Department of Physics, University of Illinois at Urbana-Champaign, Urbana, Illinois 61801-3080, USA}
\affiliation{T.C.M. Group, Cavendish Laboratory, JJ Thomson Avenue, Cambridge CB3 0HE, United Kingdom}
\author{Oliver Hart}
\affiliation{Department of Physics and Center for Theory of Quantum Matter, University of Colorado, Boulder, CO 80309 USA}
\affiliation{T.C.M. Group, Cavendish Laboratory, JJ Thomson Avenue, Cambridge CB3 0HE, United Kingdom}
\author{Cecilie Glittum}
\affiliation{T.C.M. Group, Cavendish Laboratory, JJ Thomson Avenue, Cambridge CB3 0HE, United Kingdom}
\author{Claudio Castelnovo}
\affiliation{T.C.M. Group, Cavendish Laboratory, JJ Thomson Avenue, Cambridge CB3 0HE, United Kingdom}

\begin{abstract}
\setstretch{1.1}
We analyze the out-of-equilibrium dynamics of a quantum particle coupled to local magnetic degrees of freedom that undergo a classical phase transition. Specifically, we consider a two-dimensional tight-binding model that interacts with a background of classical spins in thermal equilibrium, which are subject to Ising interactions and act as emergent, correlated disorder for the quantum particle.
Particular attention is devoted to temperatures close to the ferromagnet-to-paramagnet transition. To capture the salient features of the classical transition, namely the effects of long-range correlations, we focus on the strong coupling limit, in which the model can be mapped onto a quantum percolation problem on spin clusters generated by the Ising model. By inspecting several dynamical probes such as energy level statistics, inverse participation ratios, and wave-packet dynamics, we provide evidence that the classical phase transition might induce a delocalization-localization transition in the quantum system at certain energies. We also identify further important features due to the presence of Ising correlations, such as the suppression of compact localized eigenstates.
\end{abstract}
\maketitle
\begin{figure}[h!]
\includegraphics[width=0.84\columnwidth]{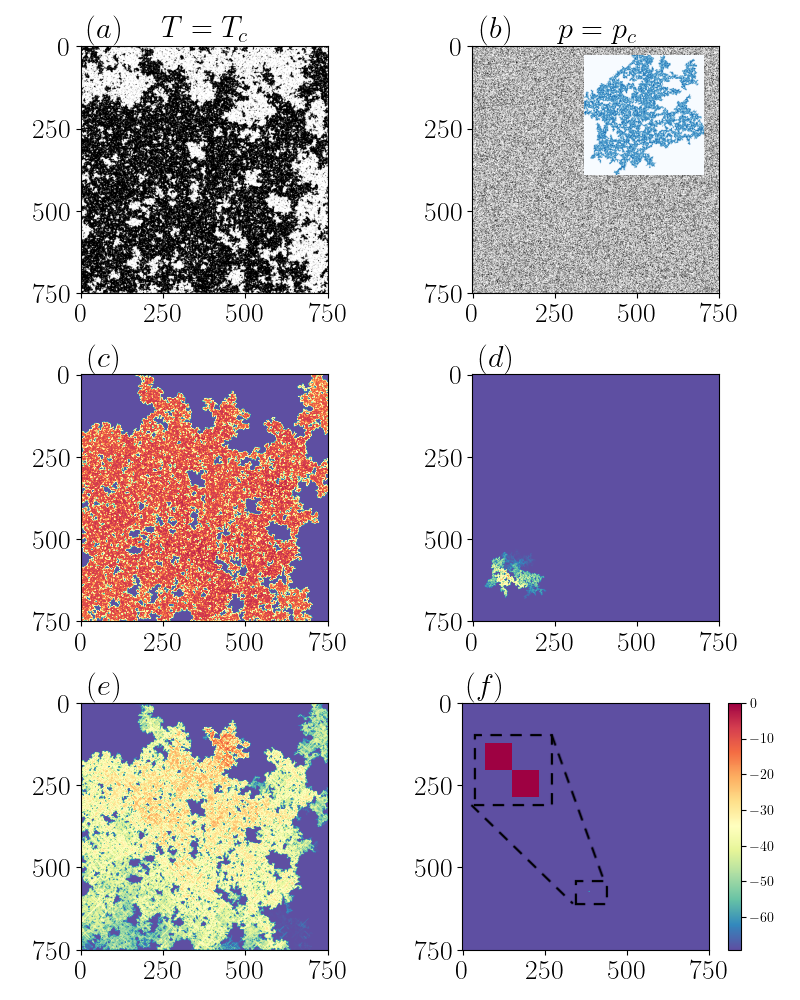}
\caption{\textbf{Spin configurations at criticality and corresponding wave function profiles}. Spin configurations for a $2d$ Ising model at the critical temperature $T = T_c$ (a), and for a classical site percolation problem at the percolation transition $p=p_c$ (b). The inset in (b) shows the largest connected cluster. (c)--(f): Amplitudes of the eigenfunctions, $\log( |\psi_{\boldsymbol{x}} (E)|^2 /\max_{\boldsymbol{x}}{|\psi_{\boldsymbol{x}} (E)|^2} )$, of the tight-binding model defined on the largest cluster of the spin configurations in (a) and (b). The left panels are for the Ising (correlated) case and the right ones for the percolation (uncorrelated) case. In (c) and in (d) the energy of the wavefunction is $E\approx 0.8$, while in (e) and (f) $E\approx 0$. In panel (f) the magnified box shows that the wave function is entirely localized on exactly two sites.
}
\label{fig:Profile}
\end{figure}
%
\section{Introduction}
After more than fifty years since its discovery, Anderson localization~\cite{Anderson58} remains a fascinating and active front of research. Anderson localization is a wave interference phenomenon in which transport in a non-interacting system can be suppressed by the presence of quenched disorder~\cite{Anderson58, Kramer1993,Evers08}. In one-dimensional systems~\footnote{Specifically, those described by short-range Hamiltonians. For a counterexample where the Hamiltonian is instead composed of long-range terms, see e.g., Ref.~\onlinecite{Mirlin_96,Evers08}.}, any amount of uncorrelated disorder is enough to exponentially localize all single-particle eigenstates, thereby generating a perfect insulator~\cite{Twose61,Goldshtein1977}. Two-dimensional systems are special: localization is weak and occurs over lengths that are exponentially large in the mean free path~\cite{Wegner1976,Licci79,Lee81}. In higher dimensions, $d>2$, it is well established that a metal--insulator transition exists, separating an extended (ergodic) phase at weak disorder from a localized one at strong disorder~\cite{Anderson1973, Logan_90, Mirlin91,tarquini_2016}.

Other systems in which Anderson localization plays a central role are ones where disorder is due to the  geometry of the system, e.g., random graphs~\cite{Grebenkov,Harry82,Schubert2009}. In general, in these systems, particles hop unimpeded but scatter from the system's rough edges, thus producing wave interference and potentially suppressing transport. Structural disorder can be found in many contexts, ranging from biological to quantum spin systems~\cite{Igarashi95,Essler07,Gong10,Felix2013,Agrawal19,Chepiga19,Nata20}. Recently, for example, the physics of systems with geometrical disorder was applied to study the dynamical properties and localization of quasiparticles in dimer, vertex, and ice models, where local constraints force the quasiparticles to move on random structures~\cite{Ollie_Claudio20,Ollie20,Stern21}.

An outstanding puzzle involving structural disorder is \textit{quantum percolation}~\cite{Schubert2009, Nakanishi2009}. In a percolation problem, one asks if a particle can propagate unboundedly on a lattice (say a regular lattice in $d$ dimensions, $\mathbb{Z}^d$) where sites have been removed at random with probability $1-p$ ($0<p<1$). In the classical case, one finds a well-defined transition at the so-called percolation threshold $p_c$: For $p>p_c$, an infinite connected cluster exists (spanning cluster), while for $p<p_c$ the system fragments into small finite clusters~\cite{Coniglio_1977, Domb_1977,Coniglio_1979, Essam_1980,grimmett1999percolation,Peliti_2011}. In the quantum realm, the presence of a spanning cluster is not sufficient to guarantee the existence of extended states, since the geometrical disorder produced by the irregular shape of the cluster might induce Anderson localization. The possibility of a quantum percolation transition at some threshold $p_Q \ge p_c$ has been extensively investigated. In two dimensions ($d=2$), its existence is still under debate~\cite{Sou91,Sriva84,Berkovits96, Naza02,   Islam08, Schubert08, Gong09,  thomas2017,Shapir1982, Shapir1982, Dillon2014}. Several numerical works have claimed to show that all the eigenstates are exponentially localized for any $p < 1$, in agreement with the one-parameter scaling theory~\cite{Sou91, MOOKERJEE95,Shapir1982}. However, other numerical works brought this conclusion into question by presenting evidence in favour of a quantum percolation transition at some $p_c \leq p_Q < 1$~\cite{Sriva84, Islam08, Schubert08, Gong09, Naza02, thomas2017, Dillon2014}. 

An interesting mechanism that can alter the nature of the eigenstates is the introduction of correlations in the disorder~\cite{Ziman82,Gri88,Phill90, Phil94, Delo98,Izra99,Shima04,Kuhl08, IZRAILEV2012125, Croy2011, GDT16}. Even in one dimension, where the system is localized for any amount of uncorrelated disorder, the presence of correlations can either partially or completely destroy localization. For instance, dimerization of the onsite potential in a $d=1$ tight-binding model 
(when the potential appears in identical pairs on adjacent sites) 
has been shown to generate extended states~\cite{Phill90, Phil94}. This demonstrates how short-range correlations are already sufficient to modify the localized nature of the system~\cite{IZRAILEV2012125}. The presence of correlations in the disorder is not just a natural and interesting question \textit{per se}, but has found important applications in disordered conducting polymers~\cite{ Peng1992, Holste2001,Krokhim_2009}, graphene~\cite{Li_2011}, quantum Hall wires~\cite{Tohru2007}, topological phases~\cite{Girschik13, Behrends2017, Mascot19}, and trapped-ion experiments~\cite{Garttner21,Martin21}.

In our work, we investigate the effects of correlations on the out-of-equilibrium dynamical properties of a tight-binding model in two dimensions with Ising-like disorder, i.e., taking discrete values $\pm W$. In App.~\ref{app:potts} we also investigate and contrast the case of $q$-state Potts disorder. The correlations are due to interactions between the disorder degrees of freedom in thermodynamic equilibrium at some finite temperature $T$. Of particular interest will be the case where the interactions cause the disorder to undergo a phase transition, at a critical temperature $T_c$ that separates a ferromagnetic phase from a paramagnetic one.

At the critical temperature the correlation length of the spin degrees of freedom diverges and we expect the largest connected cluster to dominate the behavior of the system. Thus, we focus only on the behavior of a tight-binding model defined on the largest spin cluster. 
This is in the spirit of the strong disorder limit ($W \to \infty$), in which the quantum tunneling between regions with different onsite energies can be neglected. In this limit, the system is trivially localized in the paramagnetic phase ($T>T_c$), since the spin configurations fragment into a distribution of clusters with finite size~\cite{Stella89}. However, in the ferromagnetic phase ($T<T_c$) a percolating cluster
exists~\cite{Hu84,Paniagua1997, McCoyWu_2013}.
As previously mentioned, the extensive size of the percolating cluster does not guarantee the presence of extended states and ergodicity; its `rough' edges can induce Anderson localization, resulting in the absence of diffusion.

Our results uncover a rich and interesting phenomenology that quantifies the role of correlations in the disorder. Fig.~\ref{fig:Profile} shows the typical behavior of the quantum percolation problem for the Ising case (left panels) and for the uncorrelated case (right panels). In particular, panels (a) and (b) show the spin configuration for the Ising case close to the Ising transition ($T = T_c)$ and for the uncorrelated case close to the $2d$ site percolation transition ($p = p_c$), respectively. The amplitudes of the eigenstates for the tight-binding model defined on the largest percolating cluster are shown in Figs.~\ref{fig:Profile} (c)--(f). For the case of Ising correlated disorder, we can have both uniformly spread wavefunctions, see Fig.~\ref{fig:Profile} (c), or localized wavefunctions, see Fig.~\ref{fig:Profile} (e), depending on the energy. However, in the uncorrelated case, only localized wavefunctions are present, including those with strictly vanishing localization length, dubbed \emph{compact localized states} (CLS)~\cite{Kirkpa72, Ollie20}. By inspecting several dynamical probes, we observe an important qualitative change in behavior between $T < T_c$ and $T > T_c$. The latter is consistent with the particle being localized. The former, on the other hand, exhibits dynamics akin to (quantum) diffusion, with the diffusive behaviour becoming progressively more anomalous as the critical temperature is approached from below (see e.g., Fig.~\ref{fig:Dynamics}). Using finite size scaling analysis, we provide evidence in support of the existence of a possible quantum percolation transition that coincides with the classical critical temperature.

Our model is closely related to the Falikov--Kimball model~\cite{Hubbard1963,Falicov69}, which is one of the paradigmatic models used to describe strongly correlated electrons~\cite{Hubbard1963,Falicov69,Brandt1989,Freericks03,Antipov14, Antipov16, Herrmann18}. In the Falikov--Kimball model, electrons interact with classical Ising background fields, which 
generate an effective disorder potential~\cite{ Falicov69,Antipov16,Knolle21}. Despite its translationally invariant nature, the system can exhibit Anderson localization. Recently, this connection has received a growing interest in the context of disorder-free localization~\cite{Adam2017,Adam2018,Brenes18,smith2019thesis,Knolle21} and our work provides a further example along this line of research. 

The rest of the paper is organized as follows. In Sec.~\ref{sec:model} we introduce the model and discuss its strong disorder limit and its connection to the quantum percolation problem. The methods and probes used to investigate the out-of-equilibrium dynamics are discussed in Sec.~\ref{sec:method}. The main results are presented in Sec.~\ref{sec:results}. Namely, in Sec.~\ref{sec:Static} we study the eigenvalue and eigenstate properties of the system in the limit of strong disorder as a function of the Ising temperature. We discuss possible scenarios and, in particular, we show that our results are consistent with the existence of a correlation-driven quantum percolation transition. The finite-time quantum evolution is described in Sec.~\ref{sec:Dynamics}. By detecting the spread of a particle initially localized on a single lattice site, we show that the system exhibits nontrivial dynamics for temperatures belonging to the ferromagnetic phase of the Ising model. Finally, in Sec.~\ref{sec:conclusion}, we present our conclusions and outlook.
%
%

\section{Model} \label{sec:model}

We study a two-dimensional model representing 
a tight-binding particle coupled to classical spins, described by the Hamiltonian 
\begin{equation}\label{eq:Hamiltonian}
    \hat{H}_W =
    -t \sum_{\langle \vec{x}, \vec{y} \rangle} |\vec{x} \rangle \langle \vec{y} |
    +
    W \sum_{\vec{x}} \sigma_{\vec{x}} |\vec{x}\rangle \langle \vec{x} |
    \, ,    
\end{equation}
represented in the site basis $\{ |\bm x\rangle \}$ of a square lattice of linear size $L$. The first sum runs over pairs of nearest-neighbor lattice sites, $\langle \vec{x}, \vec{y} \rangle$, and $t$ and $W$ are the hopping amplitude and onsite coupling strength, respectively. Without loss of generality, we shall set $t=1/2$ throughout our work. The onsite energies are parametrised by classical spins, $\{\sigma_{\vec{x}}\}$, which assume the values $\sigma_{\bm x}=\pm 1$, and are drawn from the Boltzmann probability distribution of a $2d$ classical Ising model at temperature $T$. 
The probability to be in the configuration $\vec{\sigma} = \{\sigma_{\vec{x}}\}$ is given by $P(\vec{\sigma}) = \frac{e^{-H_\text{I}(\vec{\sigma})/T}}{Z}$, where $H_\text{I}(\vec{\sigma}) = -\sum_{\langle \vec{x}, \vec{y} \rangle} \sigma_{\vec{x}} \sigma_{\vec{y}}$ is the classical Ising Hamiltonian and $Z = \sum_{ \bm \sigma} e^{-H_\text{I}(\bm \sigma)/T}$ is its partition function at temperature $T$.
The classical spin configurations 
are obtained using standard Monte Carlo simulations with the Swendsen--Wang  algorithm (which utilises cluster updates)~\cite{Swendsen87}. 
In App.~\ref{app:potts} we consider, in an equivalent way, the case where the classical degrees of freedom take three values, $\sigma_{\bf{x}} \in \{-1,0,1\}$, and interact via a classical Potts Hamiltonian. 

The Ising model is a cornerstone of statistical mechanics; in $2d$ it exhibits a symmetry breaking phase transition at $T_c = \frac{2}{\log{(1+\sqrt{2})}} \approx 2.269$, separating the ferromagnetic ($T<T_c$) and the paramagnetic ($T>T_c$) phases~\cite{Ising1925, Onsager44, baxter2007exactly}. These two classical phases can be detected using a local order parameter
\begin{equation}\label{eq:magne}
    M(T) = \frac{1}{L^2}\sum_{\vec{x}} \sigma_{\vec{x}}
    \, ,
\end{equation}
namely the magnetization per site, which is different from zero in the ferromagnetic phase and vanishes in the paramagnetic one. In $d=2$, the magnetization can be expressed in closed form~\cite{Potts1963, McCoyWu_2013} as 
\begin{equation}
M(T) = 
\begin{cases}
    \left (1- \frac{1}{\sinh{\frac{2}{T}}}\right )^{1/8} &\text{for } T< T_c, \\
    0                                                     &\text{otherwise} . 
\end{cases}
\end{equation}
Close to the critical point, on the ferromagnetic side of the transition, $M \sim (T_c-T)^{\beta}$, which defines the critical exponent $\beta = 1/8$. 
The correlation length $\xi$ is defined through the asymptotic behaviour of the two-point spin correlation function, and diverges as $\xi \sim |T-T_c|^{-\nu}$ at the critical point with $\nu = 1$~\cite{Onsager44, baxter2007exactly, Croy2011}.

A few considerations are in order. The temperature $T$ tunes the distribution of the disorder and thence its correlations.
At $T=0$, the magnetization $M(T=0) = \pm 1$ and the system is clean; the eigenstates of $\hat{H}_W$ in Eq.~\eqref{eq:Hamiltonian} are then extended plane waves.

As the temperature is increased, thermal fluctuations induce small clusters of classical spins with a sign that opposes the bulk magnetization.
These thermal fluctuations play the role of disorder in the system, which is correlated with a typical length scale given by $\xi$. In particular, at the Ising critical point, $\xi$ diverges and the disorder becomes scale-free, meaning that the correlation functions decay algebraically with the distance. Importantly, in the ferromagnetic phase ($T<T_c$) a spanning cluster exists composed of spins with the same sign, while in the paramagnetic phase the largest cluster is finite~\cite{Paniagua1997,Hu84, Croy2011}.

The study of dynamical properties of $\hat{H}_W$ in Eq.~\eqref{eq:Hamiltonian} as a function of the spin temperature $T$ is the main aim of our work. In $2d$ this is known to be a challenging task, since the localization length is believed to be exponentially large in the mean-free path~\cite{Licci79, Evers08}. To be able to study larger system sizes and consequently perform a more accurate scaling analysis, and to better capture the long-range correlations at the Ising critical point, we mainly focus our attention on the strong disorder limit ($W \rightarrow \infty$). In this limit, quantum tunneling between regions of the lattice with different onsite potential is suppressed and we approximate the behavior of $\hat{H}_W$ by restricting it to the largest cluster composed of classical spins possessing the same value. Close to the critical point, the largest cluster is generally expected to dominate the behaviour of the system. Thus, we end up with a tight-binding model defined on a highly irregular lattice: 
\begin{equation}\label{eq:Hamiltonian1}
    \hat{H}_\infty =
    -\frac{1}{2} \sum_{{\langle \vec{x}, \vec{y} \rangle \in  C } } |\vec{x}\rangle \langle \vec{y} |
    \, ,  
\end{equation}
where the sum runs over the nearest-neighbor lattice sites that belong to the largest connected cluster, $C$, of the classical $2d$ Ising model (see Fig.~\ref{fig:Profile}). 
As already mentioned in the introduction, the uncorrelated case, in which $\sigma_{\vec{x}} = 1$ with probability $p$ and $\sigma_{\vec{x}} = -1$ with probability $1-p$, has been extensively studied. The classical case on a square lattice has a site percolation transition at $p_c\approx 0.5927$~\cite{grimmett1999percolation,Peliti_2011}. The existence of a metal--insulator transition for the quantum case is still subject to controversy. In agreement with the one-parameter scaling hypothesis, several works provide evidence that all the eigenstates are localized for $p>p_c$~\cite{Shapir1982,Sou91,MOOKERJEE95}, while others show the possible existence of extended states at some specific energies, at least for $p \geq p_Q$ for some $p_Q \geq p_c$~\cite{Sriva84, Naza02, Islam08, Schubert08, Gong09, thomas2017, Qi2019}. To compare the Ising correlated model in our work with the uncorrelated case, we parametrize the probability $p$ as 
\begin{equation}\label{eq:probability}
    p(T) = \frac{M(T)+1}{2}
    \, ,
\end{equation}
where $M(T)$ is the magnetization per site of the $2d$ Ising model defined in Eq.~\eqref{eq:magne}. Thus, the spin configurations for the uncorrelated case will have the same magnetization as the Ising model. This procedure is equivalent to drawing spin configurations $\vec{\sigma}$ from the Boltzmann distribution and subsequently destroying all spatial correlations by randomly permuting the spins.

\section{Methods} \label{sec:method}

With the aim to understand the localization properties, we use several probes that address different facets of the system in question. 

First, we consider the level spacing statistics of the eigenenergies. This is quantified by the so-called $r$-value~\cite{Oga07,Atas13}, which measures the degree of repulsion between two adjacent energy levels,
\begin{equation}\label{eq:r_value}
    r = \langle \min(\delta_n / \delta_{n+1}, \delta_{n+1} / \delta_n ) \rangle 
    \, ,
\end{equation}
where $\delta_n$ is the level spacing between two adjacent energies and $\langle \, \cdot \, \rangle$ refers to the combined disorder and spectral averaging.
In an ergodic phase, the $r$-value is expected to be $r_{\text{GOE}} \approx 0.531$~\cite{Oga07,Atas13}, which is the same as that of a random matrix.
Conversely, in a localized phase, energy levels are uncorrelated and the $r$-value is $r_{\text{Poisson}} = 2\log{2}-1 \approx 0.386$~\cite{Oga07,Atas13}.

The second marker that we use to discern an extended phase from a localized one is the generalized inverse participation ratio, $\IPR_q$~\cite{Evers08}, 
\begin{equation}
    \IPR_q(E) = \sum_{\vec{x}} |\langle \vec{x}| E\rangle|^{2q} 
    \, ,
\end{equation}
where $|E\rangle$ is an eigenstate of $\hat{H}_\infty$ at energy $E$.
We focus our attention on two values of $q$: $q=1/2$ and $q=2$. In an ergodic phase, $\IPR_{1/2}\sim L$ and $\IPR_{2} \sim L^{-2}$, while in a localized phase $\IPR_{1/2,2} \sim {\cal O}(L^0)$. To better quantify the scaling of the $\IPR_q$ with system size, we define the multifractal exponent
\begin{equation}\label{eq:D_multi_exp}
    D_q = \frac{1}{1-q}\frac{\langle \log{\IPR_q} \rangle}{\log{L^2}}
    \, ,
\end{equation}
which takes the value $D_q=1$ in an ergodic phase and $D_q=0$ in a localized phase. Generally, $D_q$ can assume intermediate values, $0 < D_q < 1$, in which case we say that the state presents multifractality. Multifractal states are characterized by strong fluctuations of their amplitude in space, such as those found exactly at the Anderson transition~\cite{Evers08}. 

Finally, we study the out-of-equilibrium dynamics by analyzing the spread of a wave packet initially localized at $\vec{x}_0$ (i.e., in the state $| \vec{x}_0 \rangle$). The subsequent spread of the wave packet is analyzed using two different quantities: the survival probability, 
\begin{equation}
    R(t) =
    \langle |\langle \vec{x}_0 |e^{-i \hat{H}_\infty t} |\vec{x}_0 \rangle |^2 \rangle
    \, ,
    \label{eq:return}
\end{equation}
and the mean-square displacement, 
\begin{equation}
    \langle \Delta X^2(t) \rangle = \left\langle\sum_{\vec{x}} d(\vec{x}, \vec{x}_0)^2 |
    \langle \vec{x} |e^{-i \hat{H}_\infty t} |\vec{x}_0\rangle |^2 \right\rangle
    \, ,
    \label{eq:x_2}
\end{equation}
where $d(\vec{x}, \vec{x}_0)$ is the euclidean distance between the two lattice points $\vec{x}_0$ and $\vec{x}$ belonging to $\mathbb{Z}^2$.
In Eqs.~\eqref{eq:return} and~\eqref{eq:x_2} the expectation value $\langle \, \cdot \, \rangle$ includes an average over the initial site $|\vec{x}_0\rangle$ in addition to the average over disorder (i.e., over Ising configurations). 

The return probability $R(t)$ is a local probe, and is therefore sensitive to localization. Indeed, if the system has a nonzero density of localized eigenstates, then the survival probability converges at asymptotically long times to a system-size-independent value given by
\begin{equation}
    \lim_{\tau\rightarrow \infty} \frac{1}{\tau}\int_0^{\tau} dt \, R(t)
    =
    \sum_{E} \left\langle |\langle \vec{x}_0| P_{E}  | \vec{x}_0 \rangle |^2 \right\rangle
    \, ,
\end{equation}
where $P_E$ is the projector onto eigentates at energy $E$. 
On the other hand, the mean-square displacement $\langle \Delta X^2(t) \rangle$ is a \emph{global} measure and hence its behaviour reflects the existence of extended states. 

We are also interested in the dynamics at finite time scales. In order to characterize the type of propagation, we define the dynamical exponent
\begin{equation}
    \label{eq:alpha}
    \alpha(t) = \frac{d \log{  \langle \Delta X^2(t) \rangle} }{d \log{t}}
    \, . 
\end{equation}  
The propagation of the wave packet is ballistic for $\alpha(t) = 2$, diffusive for $\alpha(t) = 1$ and anomalous (subdiffusive) if $0<\alpha(t)<1$. 

\section{\label{sec:results}
Results}

\begin{figure}[h!]
    \includegraphics[width=1.\linewidth]{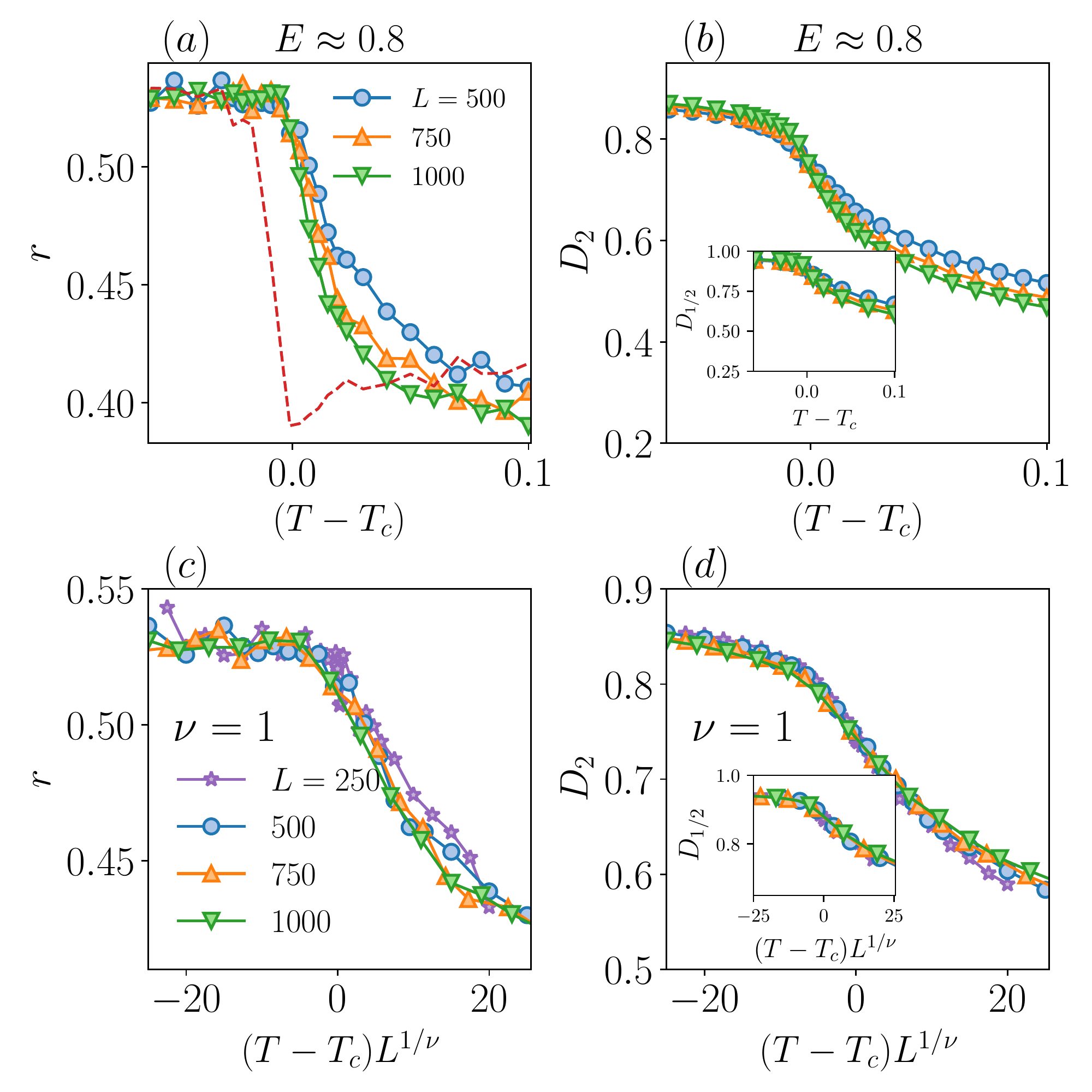}
    \caption{\textbf{Analysis of eigenvalue and eigenstate properties at $E\approx 0.8$}. (a): $r$-value statistics as a function of temperature $T$ for several system sizes $L$. The dashed line in (a) is for uncorrelated disorder with the largest system size. (b): Multifractal exponent $D_2$ as a function of temperature $T$ for the same parameters as panel (a). The inset in (b) shows $D_{1/2}$ as a function of $T$. (c): Finite size collapse of the $r$-value statistics (a). (d): Finite size collapse of $D_2$ (b) The inset shows the finite size collapse for the $D_{1/2}$ multifractal exponent. Both panels (c) and (d) use the Ising critical exponent $\nu = 1$.}
    \label{fig:Eigen}
\end{figure}

\subsection{Spectral and eigenstate properties} \label{sec:Static}

In this section, we analyze the properties of the eigenvalues and eigenstates of $\hat{H}_\infty$ in Eq.~\eqref{eq:Hamiltonian1}. In order to understand the localization properties of the system, we focus on the $r$-value in Eq.~\eqref{eq:r_value} and the $D_q$ exponents in Eq.~\eqref{eq:D_multi_exp}. The results presented in this section were obtained using shift-invert diagonalization techniques, computing $32$ eigenstates and eigenvalues at a specific target energy $E$~\cite{Francesca2018}.

Before starting, it is important to point out that systems with geometrical disorder have the peculiarity that states with the smallest localization length are found in the \emph{middle} of the spectrum ($E=0$ in our case), in contrast to systems subject to diagonal disorder (see e.g., Refs.~\onlinecite{Nakanishi2009, Schubert2009, Ollie20}). Thus, we focus our investigation on two energy values, one far away from and one at the band center. Fig.~\ref{fig:Eigen} and~\ref{fig:E_0} show the $r$-value, $D_{1/2}$, and $D_{2}$, for energies $E\approx 0.8$ and $E\approx 0$, respectively~\footnote{We checked the behavior of the system also at other energies away from the central band, finding similar results. See, e.g., App.~\ref{app:ising}}.

\subsubsection{$E\approx 0.8$}
Let us consider the data at $E\approx 0.8$. Panels (a) and (b) in Fig.~\ref{fig:Eigen} are dedicated to the $r$-value and to the two fractal exponents $D_{1/2, 2}$, respectively, as a function of $T-T_c$. In Fig.~\ref{fig:Eigen} (a) the $r$-value reaches the GOE result for $T<T_c$, while for $T>T_c$ the curves deviate from the ergodic behaviour showing a trend towards the Poissonian value with increasing $L$. 

Figure~\ref{fig:Eigen} (a) also shows the $r$-value of the uncorrelated model defined in Sec.~\ref{sec:model} for the largest system size, $L=1000$. First, we notice that for $T-T_c \lesssim -0.025$ the $r$-statistics approaches the GOE value. Whether this behavior is due to the existence of a metal-insulator transition for the uncorrelated case, as asserted in several works~\cite{Sriva84, Naza02, Islam08, Schubert08, Nakanishi2009, Schubert2009, Gong09, thomas2017, Qi2019}, or whether it merely indicates a finite size crossover in which the localization length becomes comparable to the system size $L$, is beyond the scope of our work. However, it is interesting to observe that at $T\approx T_c - 0.025$ the magnetization takes the value $M\approx 0.64$,
corresponding via Eq.~\eqref{eq:probability} to an uncorrelated probability $p \simeq 0.82$,
which is notably larger than the classical percolation threshold $p_c \approx 0.59$.
Thus, if a delocalization transition does take place in the uncorrelated case (quantum percolation), we can at least propose a lower bound for it: $p_Q \geq 0.82 > p_c$. As a result, $T\approx T_c-0.025$ represents the temperature below which we cannot distinguish the role of Ising interactions/correlations. Due to the limitation in system sizes, we were not able to make conclusive statements at small temperatures and we focus our analysis on the regime $T-T_c \ge -0.025$.

In agreement with the $r$-value statistics, the behavior of the fractal exponents $D_{1/2, 2}$ changes around $T\approx T_c$, see Fig.~\ref{fig:Eigen} (b) and its inset. In the ferromagnetic phase the fractal exponent tends to the ergodic value ($D_q\rightarrow 1$), while in the paramagnetic phase it tends to the localized value ($D_q\rightarrow 0$) since the largest connected cluster is finite. 

It is tempting to suggest that the data are consistent with the
emergence of a metal--insulator transition at $T\approx T_c$, at least for certain energies.
In this scenario, the quantum percolation threshold coincides with the classical Ising transition.
For $T<T_c$ the wave functions are ergodic in the largest cluster, implying that $D_q \rightarrow 1$ in the thermodynamic limit ($L\rightarrow \infty)$, whereas $D_q\rightarrow 0$ trivially in the paramagnetic phase.
It would then be reasonable to expect the behaviour at the transition point to be dominated by the Ising criticality and thus the quantum and classical cases share the same critical exponents. To support this idea, we rescale the two dimensionless probes, $r$ and $D_q$, using the known classical critical exponent $\nu=1$ that governs the correlation length $\xi \sim |T-T_c|^{-\nu}$ of the $2d$ Ising transition. We find a good collapse, as shown in Fig.~\ref{fig:Eigen}~(c)-(d), which also provides further evidence in favour of a quantum percolation transition. 

However, it is important to point out that -- keeping in mind the difficulties associated with detecting localization properties in $2d$ and absent an analytical solution -- we must remain open to the possibility that the system is localized for $T<T_c$, but with a localization length much larger than the linear system sizes we are able to access numerically. In favor of a possible large localization length, we can refer to the recent work in Ref.~\onlinecite{Ollie20} where a $1d$ tight-binding model with linear offshoots whose lengths are distributed randomly (random quantum combs) has been investigated. The random quantum comb shares some similarities with our problem and it could be thought as a quasi-$1d$ version thereof, without correlations. Using analytical and numerical techniques, Ref.~\onlinecite{Ollie20} showed that the system is localized with a localization length extremely large at some energies (where for uniform hopping one finds $\xi_{\text{loc}}\sim {\cal O}(10^3)$). In this latter scenario, a finite size crossover between ergodic and localized behaviour would shift to lower and lower temperatures as the system size is increased, possibly down to $T=0$ ($p_Q \to 1$). In App.~\ref{app:mean_free}, we present a pedagogical computation, in the weak-scattering approximation~\cite{Bruus2003} $W\ll 1$ and $T\ll T_c$, which allows to estimate the mean-free path, $\ell$, as a function of temperature and disorder amplitude. This argument provides evidence that $\ell \propto \frac{(T_c/T-1)}{W^2}$ which, combined with the one-parameter scaling, implies an exponentially large localization length in $(T_c/T-1) / W^2$, for $T\ll T_c$ and $W\ll 1$. 

Although it is important to exercise caution in claiming that a transition exists, we can certainly assert that Ising correlations play an important role. Indeed, our results demonstrate that they modify substantially the localization properties -- see, e.g., Fig.~\ref{fig:Eigen} (a) -- either by increasing the localization length or by shifting to a lower temperature the putative quantum percolation transition of the equivalent uncorrelated problem with $p$ set by Eq.~\eqref{eq:probability}. 
\begin{figure}[t!]
    \includegraphics[width=1.\linewidth]{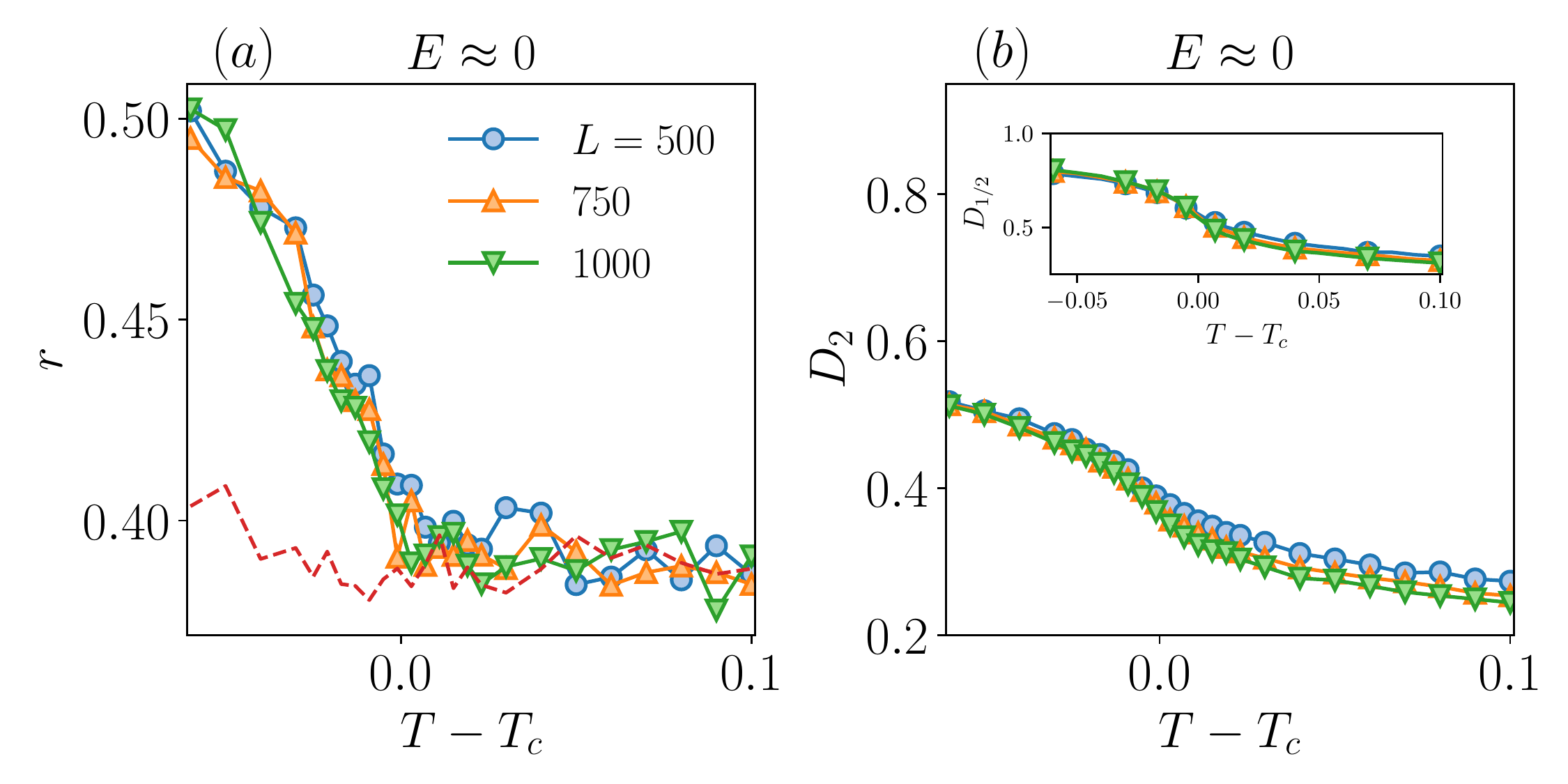}
    \caption{\textbf{Analysis of eigenvalue and eigenstate properties at $E \approx 0$}.  
     (a): $r$-value statistics as a function of temperature $T$ for several system sizes $L$. The dashed line in (a) is for uncorrelated disorder with the largest system size. (b): Multifractal exponent $D_2$ as a function of temperature $T$ for the same parameters as panel (a). The inset in (b) shows $D_{1/2}$ as a function of $T$.
    }
    \label{fig:Eigen_E_0}
\end{figure}
\begin{figure}[h!]
    \includegraphics[width=1.\linewidth]{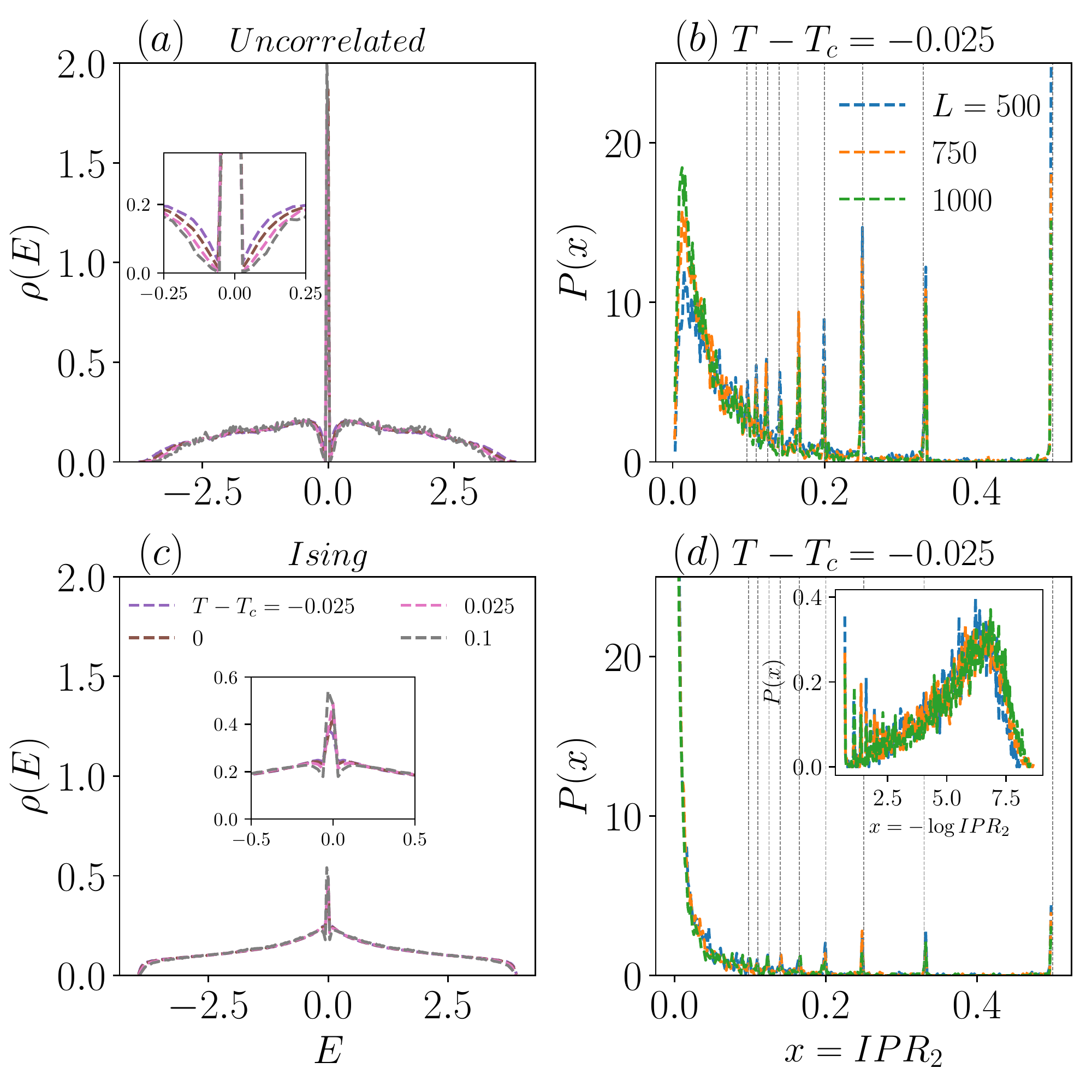}
    \caption{\textbf{Density of states and probability density of} $\mathbf{IPR}_\vec{2}$. Panels in the top row refer to the uncorrelated case, while panels in the second row refer to the Ising correlated disorder. (a), (c): Density of states $\rho(E)$, Eq.~\eqref{eq:density}, for several temperatures and $L=256$. The insets magnify the energy window around $E\approx 0$. (b), (d): Probability distribution $P(x)$ of $x=\IPR_2$ for several $L$. The vertical dashed lines are guides to the eye, indicating the values of $\IPR_2$ for the most prevalent CLS. The inset in (d) shows the probability distribution $P(x)$ of $x = -\log\IPR_2$ for the Ising case.}
    \label{fig:Density}
\end{figure}

\subsubsection{$E\approx 0$}
In this section, we consider energies belonging to the middle of the spectrum ($E\approx 0$), where we find a remarkable difference in the behaviour of the system between the Ising correlated disorder and the uncorrelated case. Fig.~\ref{fig:Eigen_E_0} is dedicated to the $r$-value and the fractal exponents $D_{1/2,2}$ at $E=0$ as a function of $T$. In contrast to the $E\approx 0.8$ data, see Fig.~\ref{fig:Eigen}, we observe a stronger trend to localization. This is in agreement with the general picture that the most localized states are near the band centre~\cite{Nakanishi2009, Schubert2009, Ollie20}. 

For the Ising interacting case in the paramagnetic phase, the $r$-value is Poissonian; in the ferromagnetic phase, it never reaches the GOE value and a trend to Poissonian is in fact visible with increasing $L$. For completeness, in Fig.~\ref{fig:Eigen_E_0} (a) we show the $r$-value for the corresponding uncorrelated case, which is Possonian. A behavior similar to the $r$-value is observed for the fractal exponent $D_2$, shown in Fig.~\ref{fig:Eigen_E_0} (b). A careful analysis shows that a slow scaling $D_2\rightarrow 0$ is present also in the ferromagnetic phase. Thus, from both probes -- the $r$-value and the $D_2$ exponent -- we would conclude that the eigenstates are localized. However, the $D_{1/2}$ exponent, which is more susceptible to extended states (see Sec.~\ref{sec:method}), saturates with $L$ to a strictly positive value for $T<T_c$. This suggests the existence of extended states or at least less localized ones. This contradictory behavior between the $r$-value, the $D_2$ and the $D_{1/2}$ exponents could be explained by the coexistence of states with different localization properties. 

In order to better understand this behavior, we take a closer look at the eigenvalues and eigenfuctions around $E=0$. Let us start with the energy spectrum, by inspecting the density of states (DOS) defined as
\begin{equation}\label{eq:density}
    \rho(E) =  \left\langle 
    \sum_i\frac{\delta(E-E_i)}{\text{dim}(\mathcal{H})} 
    \right\rangle
    \, , 
\end{equation}
where $\{E_i\}$ denotes the set of eigenvalues and $\text{dim}(\mathcal{H})$ is the dimension of the Hilbert space, which ensures the normalization $\int dE \, \rho(E) = 1$~\footnote{Notice that $\text{dim}(\mathcal{H})$ coincides with the dimension of the largest cluster.}. 

\begin{figure}[t!]
    \includegraphics[width=0.9\linewidth]{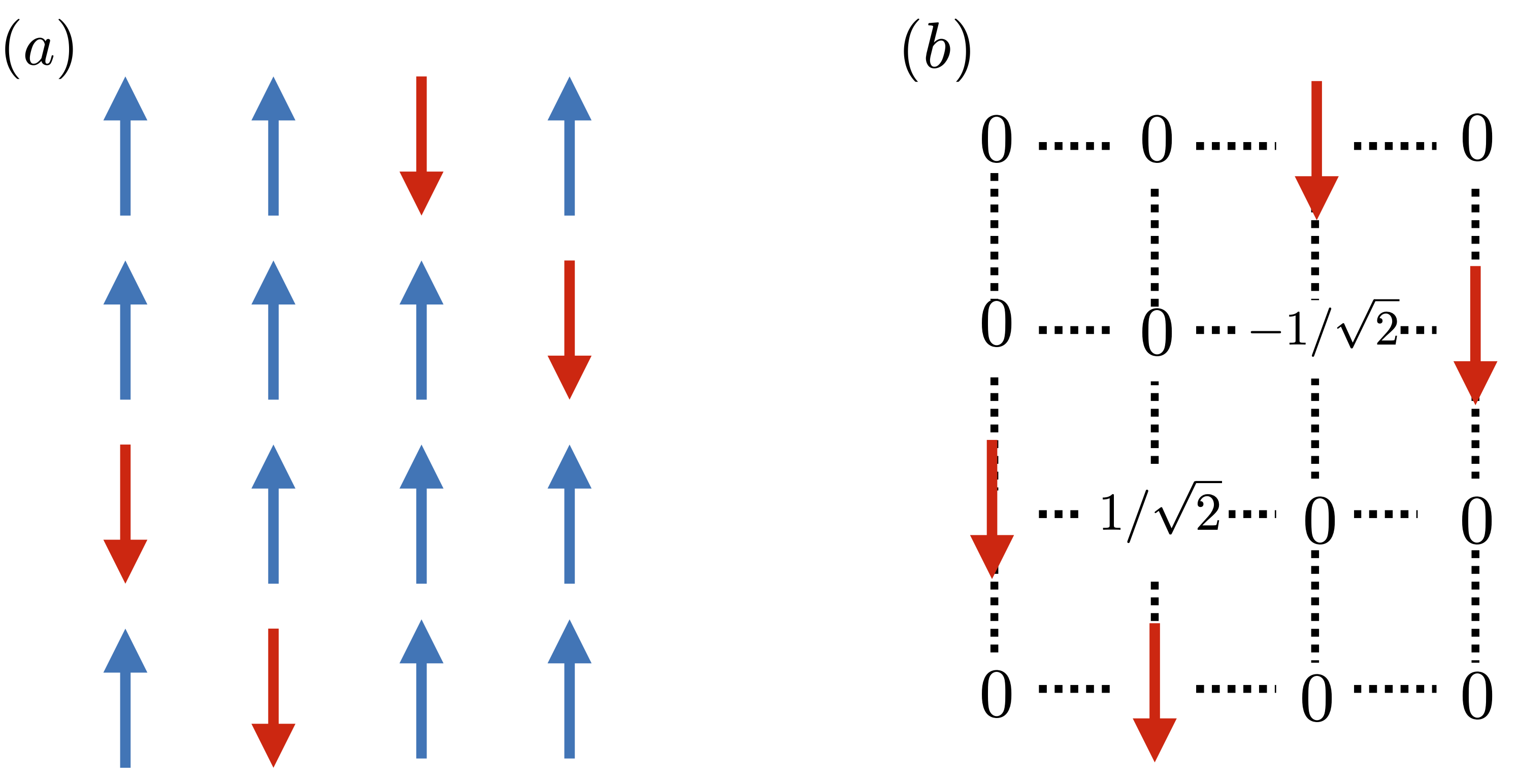}
    \caption{\textbf{Simple compact localized state.} (a): The local spin configuration that hosts a CLS zero-mode strictly localized on two sites. (b): The corresponding wave function of the $E=0$ mode.}
    \label{fig:E_0}
\end{figure}

In Figs.~\ref{fig:Density} (a) and (c) we show the DOS of the uncorrelated and of the Ising case, respectively, for several values of $T$. 
The DOS for the uncorrelated quantum percolation problem has received a lot of attention and has already been studied extensively~\cite{Chayes1986,Naumis94,Barrio1998,Numis02}. This earlier work highlighted that the DOS has a spike singularity at $E = 0$ due to the macroscopic degeneracy of special states, 
surrounded by a pseudo-gap~\cite{Chayes1986,Naumis94,Barrio1998,Numis02}, see also the inset of Fig.~\ref{fig:Density} (a).
The existence of this peculiar structure in the DOS is due to a finite density of compact localized states. The CLS are defined as eigenstates of $\hat{H}_\infty$
whose support is confined to a strictly finite number of sites, see Fig.~\ref{fig:Profile} (f). These states are typical in tight-binding models with geometrical/configurational disorder and they have been found in several models~\footnote{We note in passing that these states might persist even in the presence of local interactions, generating quantum many-body scars, i.e., atypical non-thermal eigenstates embedded in a sea of thermal states~\cite{Ollie_Claudio20,Papic2021, BernevigReview2021}.}, e.g., in random combs, random graphs, quantum spin-ice, and electronic models that host flat bands~\cite{Kirkpa72,Balents08, Green10,Flach2014,Kosior17,Carlo1,Carlo2,Carlo3,Maimaiti17,Yoshihito2020,Ollie20,Zhang20,Giedrius2020,Stern21}. In the present case, these states are generated by special \emph{local} configurations of disorder (spins)~\cite{Kirkpa72}. Fig.~\ref{fig:E_0} (a) shows an example of a local spin configuration -- see also Figs.~\ref{fig:Profile} (b) and (f) -- which hosts an $E=0$ eigenstate that is fully localized on two sites, see Fig.~\ref{fig:E_0} (b). Other spin configurations hosting different $E=0$ CLS can be found in, e.g., Refs.~\onlinecite{Kirkpa72, Schubert08, Nakanishi2009, Ollie20}. Importantly, it is easy to see that CLS of the form illustrated in Fig.~\ref{fig:E_0} appear with finite probability and, on average, their number scales with the volume of the system, $\sim L^2$. As a result, the degeneracy of states with $E=0$ is macroscopic and the DOS has a spike singularity in the middle of the spectrum~\footnote{There are also other singularities in the DOS due to the presence of CLS at different energies~\cite{Kirkpa72, Schubert08, Nakanishi2009, Ollie20}. However, the density of these CLS at $E\ne 0$ is much smaller and the resulting singularity could be difficult to see.}.

Having explained the presence of the spike singularity in the DOS, we now discuss briefly the presence of the pseudo-gap around $E=0$, Fig.~\ref{fig:Density} (a) and inset. It turns out that the pseudo-gap in the DOS is a different manifestation of the existence of a finite density of CLS in the system. In Refs.~\onlinecite{Naumis94,Numis02} the pseudo-gap is studied in detail both theoretically and numerically, and evidence is provided that the magnitude of the pseudo-gap is proportional to $1/\sqrt{f_0}$, where $f_0$ is the density of CLS at $E=0$. Therefore, the more CLS there are, the more pronounced the pseudo-gap is. Furthermore, the presence of CLS can also be observed in the probability distribution $P(x)$ of $x=\IPR_2$. $P(x)$ exhibits several spikes, highlighted by dashed lines in Fig.~\ref{fig:Density} (d), the highest of which occurs at $\IPR_2\approx 1/2$ and is due to the CLS localized on two sites, as illustrated in Fig.~\ref{fig:E_0}. Spikes at $\IPR_2 \approx 1/n$ with $n \in \mathbb{N}$ are CLS uniformly localized on $n$ sites.

Now we turn back to the Ising correlated case and we analyze its DOS shown in Fig.~\ref{fig:Density} (c) and its inset. In this case, we do not observe a pronounced pseudo-gap nor a pronounced spike at $E=0$, as seen in the inset of Fig.~\ref{fig:Density} (c). Assuming that the relation between the height of the spike with the pseudo-gap dip, $1/\sqrt{f_0}$~\cite{Naumis94,Numis02}, remains unchanged, we might conjecture that the number of CLS at $E=0$ is suppressed by the presence of Ising interactions. To verify this conjecture, we inspect the probability distribution function of $\IPR_2$ for the Ising case, Fig.~\ref{fig:Density} (d), for several system sizes $L$. The spikes at $\IPR_2 \approx 1/n$ (vertical dashed lines) -- while still present -- are now significantly smaller in height. This supports the main idea that in the case of Ising correlated disorder, CLS are rarer, although still of seemingly finite density. Indeed, most of the probability is concentrated at small values $\IPR_2\approx 10^{-3}$, e.g., $\frac{P(x\approx 10^{-3})}{P(x\approx 1/2)} \sim 10^{2}$, and a systematic shift of the highest spike to smaller values with increasing $L$ is visible. To better highlight the behavior of the most delocalized states contributing to $\IPR_2$, we consider the probability distribution $P(x)$ of $x=-\log{\IPR_2}$, shown in the inset of Fig.~\ref{fig:Density} (d). Though slow, a trend to delocalization is clear, meaning that most of the probability mass of $\IPR_2$ shifts to smaller values ($-\log\IPR_2$ to larger values).
However, the height of the spikes at $\IPR_2 \sim 1/n$ with $n \in \mathbb{N}$ is stable with $L$. This is consistent with the coexistence of more delocalized states at $\IPR_2 \ll 1$ and CLS at $E=0$. In this scenario, the fractal exponent $D_2$ tends to zero due to the presence of a finite-density of CLS, while $D_{1/2}$ can have a strictly positive value due to the most delocalized states, in agreement with the results shown in Fig.~\ref{fig:Eigen_E_0} (b) in the ferromagnetic phase. 

\subsection{Dynamical properties} \label{sec:Dynamics}

\begin{figure}[t!]
    \includegraphics[width=1.0\linewidth]{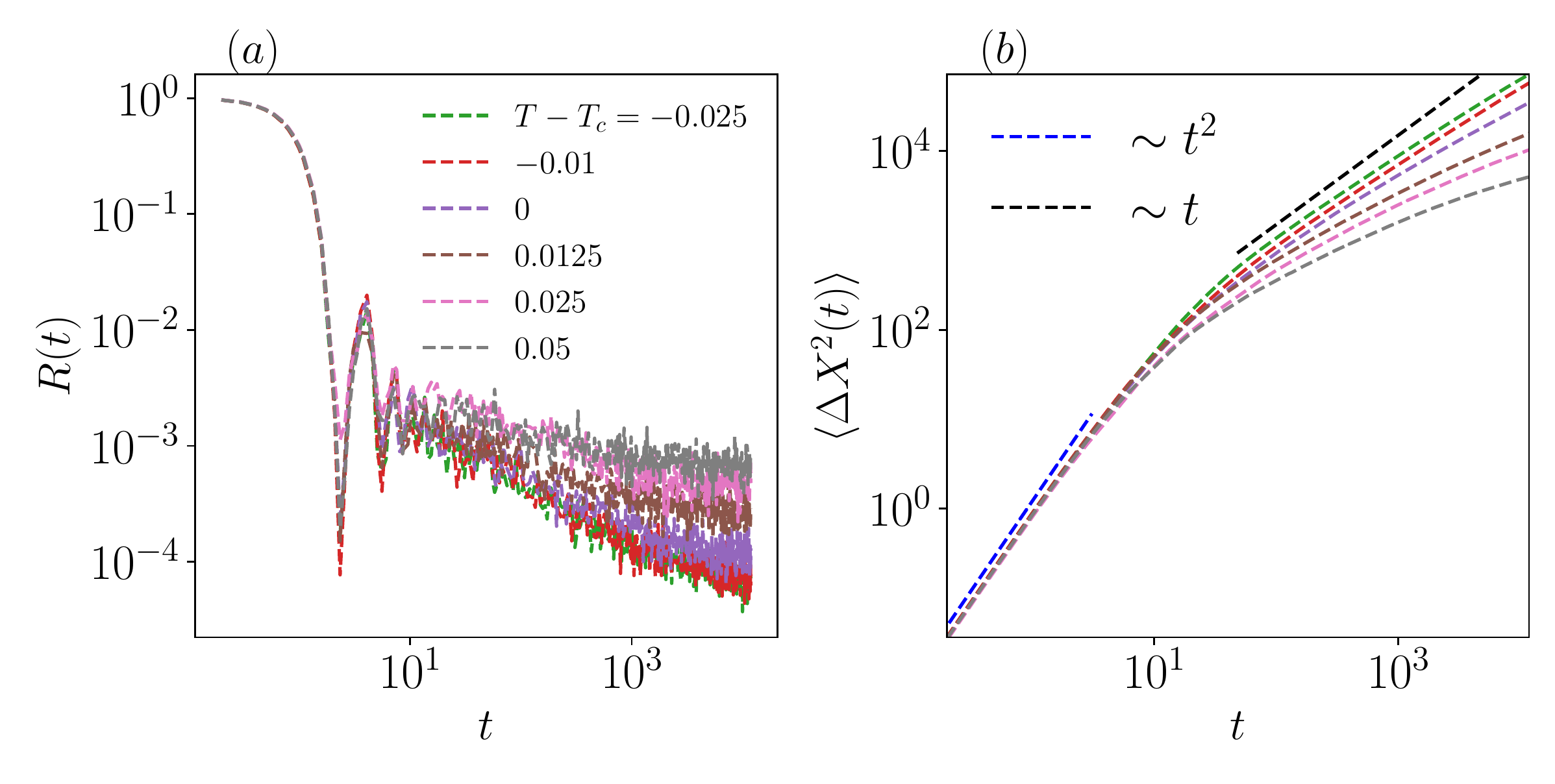}
    \caption{\textbf{Out-of-equilibrium wave packet dynamics}. (a), (b): Survival probability $R(t)$, Eq.~\eqref{eq:return}, and the expectation value of the distance $\langle \Delta X^2(t) \rangle$, Eq.~\eqref{eq:x_2}, for several $T$ after initialising the particle at a given site. The blue and the black dashed lines are guides to the eye representing ballistic and diffusive propagation, respectively. $L=1500$ in both panels.}
    \label{fig:Dynamics}
\end{figure}

Having discussed the possible existence of a metal--insulator transition that coincides with the classical critical temperature $T_c$,
we now study the finite-time dynamics of $\hat{H}_\infty$. We investigate the time evolution of a particle initialized at a given (randomly chosen) site on the cluster, and compute its probability to be on that site as a function of time, $R(t)$, Eq.~\eqref{eq:return}, as well as the expectation value of the distance from that site $\langle \Delta X^2(t) \rangle$, Eq.~\eqref{eq:x_2}. The evolution of $R(t)$ and $\langle \Delta X^2(t)\rangle$ are computed using Chebyshev integration techniques~\cite{Fehske06}, allowing us to reach large system sizes ($L_{\text{max}}= 3000$) and long times ($t_{\text{max}}\approx 10^4$). 

In particular, this section aims to shed light on the putative quantum percolation transition of $\hat{H}_\infty$ and we therefore focus our attention on temperatures close to the critical one. We report a change in the dynamical properties of the system in crossing the classical phase transition. In the paramagnetic phase, the system is localized. However, in the ferromagnetic phase ($T \leq T_c$) we see unbounded propagation for the system sizes and time scales that we are able to access numerically.

\begin{figure}[t!]
    \includegraphics[width=1.\linewidth]{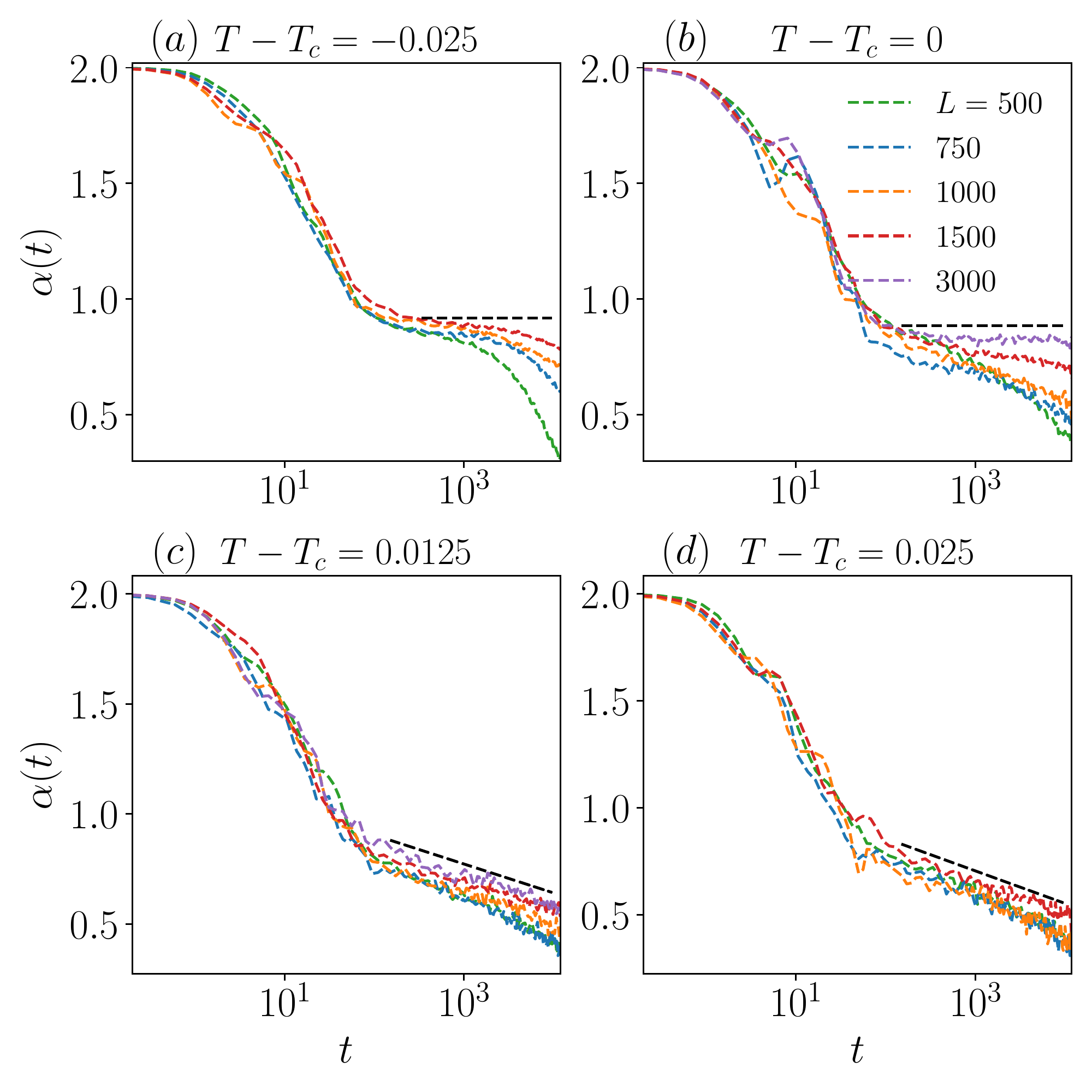}
    \caption{\textbf{Finite size effects in the exponent} $\alpha(t)$. (a)--(d) $\alpha(t)$, Eq.~\eqref{eq:alpha}, as a function of time for several $L$ and temperatures around $T_c$. The black dashed lines are guides to the eye. The largest system size ($L=3000$) is only shown in panels (b) and (c), for the temperatures closest to the critical one.}
    \label{fig:Alpha}
\end{figure}

Fig.~\ref{fig:Dynamics} shows both dynamical probes, $R(t)$ and $\langle \Delta X^2(t) \rangle$, for a range of temperatures around $T_c$, and for fixed system size $L=1500$. After a quick and temperature-independent ballistic propagation at short times, i.e., $\langle \Delta X^2(t) \rangle \sim t^{2}$, a temperature-dependent dynamics sets in.
In the paramagnetic phase ($T>T_c$), the survival probability saturates to a size-independent value $R(t \to \infty) \sim {\cal O}(L^0)$, which is a function of the temperature $T$. As expected, the saturation value of $R(t)$ decreases as the temperature is lowered, meaning that the system is more delocalized at lower temperatures. In the ferromagnetic phase, after the initial ballistic decay, $R(t)$ falls off algebraically ($R(t)\sim 1/t^{\gamma}$) up to $t \sim 10^{3}$ and then starts to saturate to a size-independent value. We note that it saturates to a value that is appreciably greater than the one imposed by the finite size nature of the system. 
It also is important to point out that the saturation of $R(t\rightarrow \infty)$ to a size-independent value is necessary but not sufficient for the system to be localized. 
Indeed, a finite \emph{density} of localized states causes $R(t)$ to saturate to a strictly positive value ($R(t\rightarrow \infty) > 0$), and therefore it does not exclude the existence of extended ones.
In the ferromagnetic phase, so long at $T>0$, we expect $R(t\rightarrow \infty)$ to saturate to a finite size-independent value in the thermodynamic limit, since the system hosts a finite density of CLS. 

The behavior of $\langle \Delta X^2(t)\rangle$ in Fig.~\ref{fig:Dynamics} (b), and the dynamical exponent $\alpha(t)$, Eq.~\eqref{eq:alpha}, in Fig.~\ref{fig:Alpha}, allow us to better investigate the presence of delocalized states for $T<T_c$. In Figs.~\ref{fig:Alpha} (a) and (b) we study the finite size effects in the time evolution of $\alpha(t)$. At short times $\alpha(t)\approx 2$ and the propagation is ballistic, as seen also in Fig.~\ref{fig:Dynamics} (b). In the ferromagnetic phase, at intermediate times, $\alpha(t)$ develops a plateau close to $\alpha \approx 1$, consistent with diffusion. Importantly, this plateau extends to larger times with increasing system size $L$, which may provide evidence that in the limit $L\rightarrow \infty$ the system is diffusive. In the paramagnetic phase instead, as expected, the dynamical exponent $\alpha(t)$ decays continuously with time, as seen in Figs.~\ref{fig:Alpha} (c) and (d), in agreement with the fact that the system is trivially localized. 

In summary, the analysis of the finite size behaviour of $R(t)$ and $\alpha(t)$ is again suggestive of a possible change in the behavior of the system when the disorder crosses the classical phase transition, and suggestive of the appearance of delocalized states for $T \lesssim T_c$ leading to a metal--insulator transition at $T \approx T_c$. 


\section{Conclusions $\&$ Outlooks} \label{sec:conclusion}

 In this work, we investigated the out-of-equilibrium dynamics of a tight-binding quantum particle coupled to interacting Ising spins in thermal equilibrium in two spatial dimensions. Of particular interest is the role played by the correlations between the spins, and by the phase transition that occurs in the Ising degrees of freedom. For temperatures close to the phase transition, long-range correlations are present and the behavior of the system is dominated by the largest connected Ising cluster. Thus, we make the working assumption of considering the model restricted to the largest cluster. This is equivalent to the strong coupling limit, in which quantum tunneling between different Ising domains is forbidden. In this limit, the model maps to a correlated quantum percolation problem, and quantum interference is produced by the highly irregular shape of the spin cluster.
 
 For $T>T_c$, the Ising model is in the paramagnetic phase and the size of the largest cluster remains finite in the thermodynamic limit. As a result, the system is trivially localized. In the ferromagnetic phase, $T<T_c$, a spanning cluster exists and we rely on exact numerical simulations to understand the localization properties of the system. By inspecting several localization markers, we observed at certain energies a strong crossover from localized to delocalized eigenstates upon crossing the Ising transition. For instance, the energy levels show repulsion and the fractal dimensions of the eigenstates tend to one in the ferromagnetic phase. The wave-packet dynamics exhibit quantum diffusion, with the diffusion becoming progressively more anomalous as the critical temperature is approached from below. By using finite size scaling analysis, we provided numerical evidence that the system might undergo a quantum percolation transition at the critical temperature of the Ising model. Throughout our work, we underlined the main differences between the Ising correlated case and the uncorrelated one. The center of the energy spectrum hosts a finite density of compacted localized states, i.e., eigenstates with strictly finite support.
These compact localized states are due to some particular local realization of disorder and they are responsible for the appearance of a pseudo-gap at the center of the density of states. We show that for the interacting case the total number of compact localized states is highly suppressed.

To summarize, we numerically investigated the dynamical properties of a particle coupled to classical Ising spins undergoing a thermal phase transition. We found an important change in the system's behavior for temperatures below and above the critical one. For temperatures belonging to the paramagnetic phase, the system is localized, while at low temperature the behavior is consistent with the existence of a delocalized phase. 

Furthermore, our work presents an example of disorder-free localization. Despite the translational invariance of the system, thermal fluctuations induce correlated disorder for the quantum particles, and localization.  

Our work paves the way for other research lines. For example, the investigation of quantum systems coupled with classical ones having first-order phase transitions, or quasi-periodic couplings, or introducing interactions between the tight-binding particles, in addition to those between the Ising degrees of freedom, are important questions that are left for future work.

\begin{acknowledgments}
We would like to thank I.M. Khaymovich and E. Fradkin for helpful discussions. GDT was supported by the Gordon and Betty Moore Foundation’s EPiQS Initiative. CG was supported by the Aker Scholarship. This work was funded in part by the Engineering and Physical Sciences Research Council (EPSRC) Grants No.~EP/P034616/1, EP/T028580/1 and~EP/V062654/1. 
\end{acknowledgments}

\appendix 

\section{$q=3$ Potts model interactions}\label{app:potts}

In this section, we inspect the behavior of the model in Eq.~\eqref{eq:Hamiltonian}, but with on-site potential $\sigma_{\vec{x}} \in \{ -1, 0, 1 \}$. The spin-configurations $\{\sigma_{\vec{x}}\}$ are drawn from the Boltzmann probability distribution of a classical $2d$ $q=3$ Potts model at temperature $T$~\cite{baxter2007exactly}. As in the main text, the classical spin configurations at a certain temperature $T$ are obtained using standard Monte Carlo simulations with the Swendsen--Wang algorithm (cluster updates)~\cite{Swendsen87}. The classical $q=3$ Potts model has a second-order phase transition at $T_c \approx 2/\log(1+\sqrt{3})$, and the correlation length diverges as $\xi \sim |T-T_c|^{-\nu}$ with $\nu \approx 0.46$~\cite{Wu_Potts_1982,Arcangelis_1986} as the critical point is approached. 

As explained in the main text, to capture the long-range correlations close to the classical phase transition, we consider the $|W| \to \infty$ limit, allowing us to focus on the tight-binding model defined on the largest connected `domain', see Eq.~\eqref{eq:Hamiltonian1}. 

\begin{figure}[t!]
    \includegraphics[width=1.04\columnwidth]{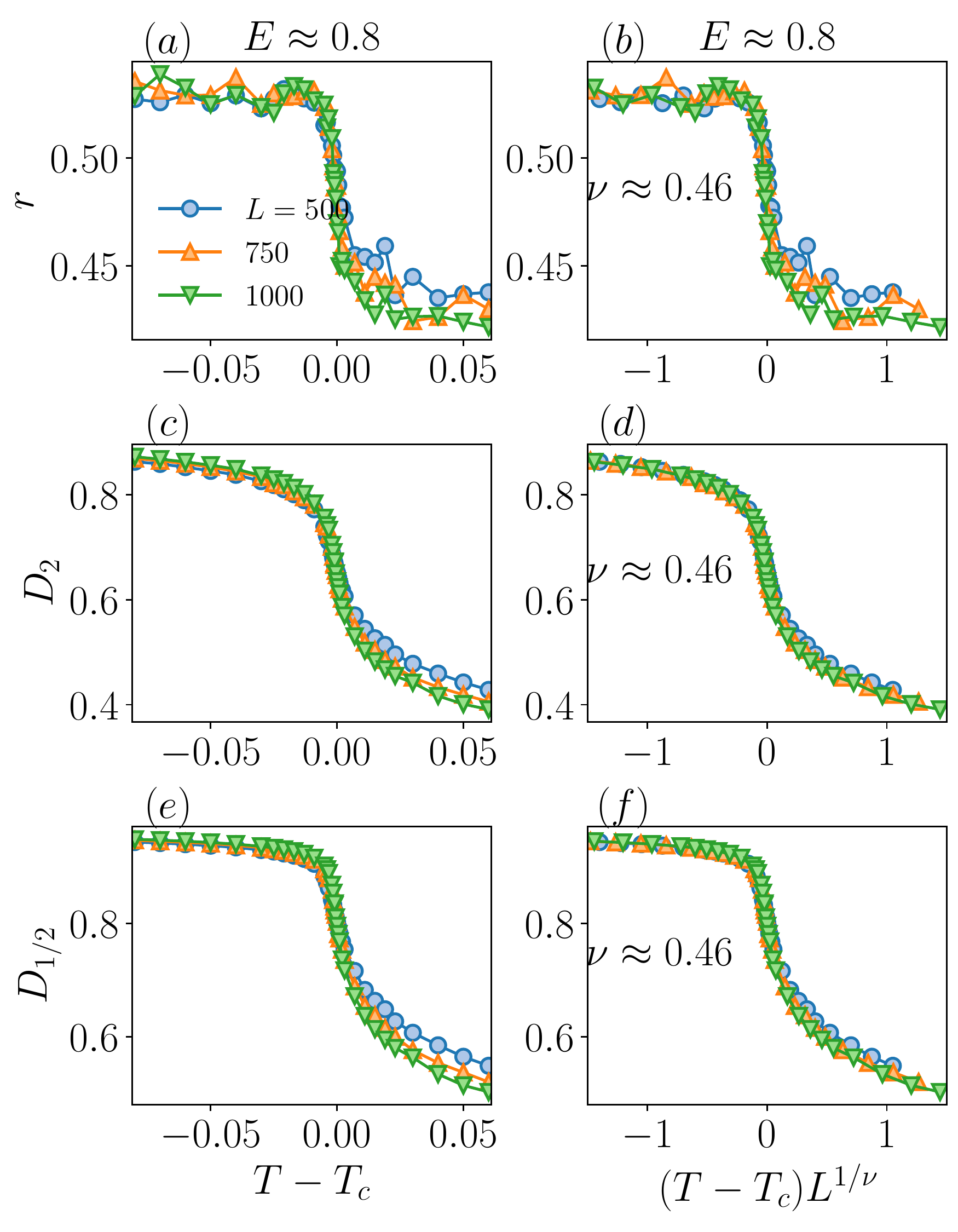}
    \caption{\textbf{Eigenvalue and eigenstate analysis for the} $q=3$ \textbf{Potts model}. (a), (c) and (e): $r$-statistics, $D_{2}$ and $D_{1/2}$, respectively, as a function of $T-T_c$. (b), (d) and (f): Collapse of the the $r$-statistics, $D_{2}$ and $D_{1/2}$ exponents, with system size.}
    \label{fig:Pott3}
\end{figure}

\begin{figure}[t!]
    \includegraphics[width=1.0\columnwidth]{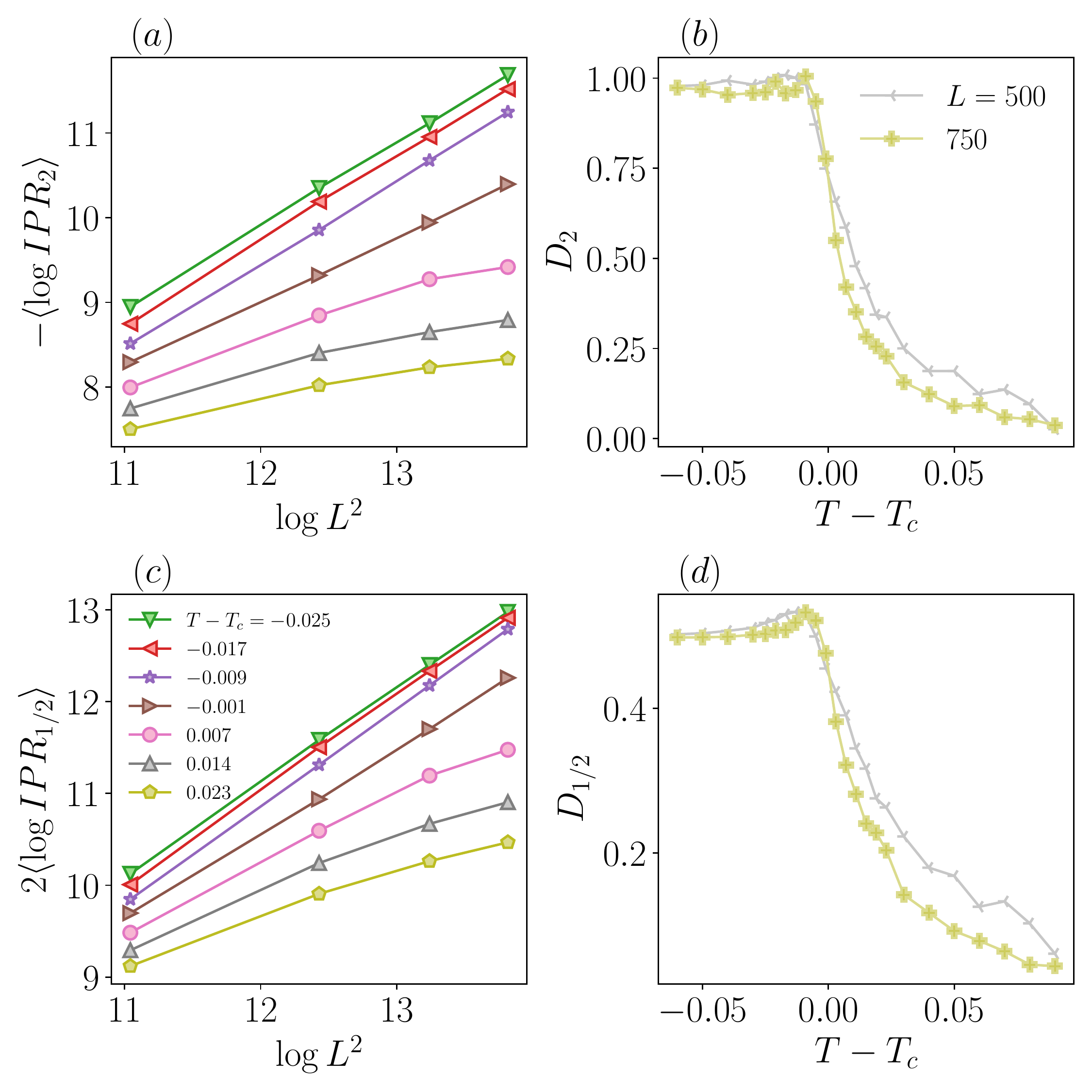}
    \caption{\textbf{$D_2$ and $D_{1/2}$ analysis for the Ising correlated case}. (a) and (c): $\IPR_{2,{1/2}}$ as a function of system size for several temperatures. (b) and (d): $D_{2,1/2}$ as a function of $T-T_c$.}
    \label{fig:Scaling1}
\end{figure}

\begin{figure}[t!]
    \includegraphics[width=1.0\columnwidth]{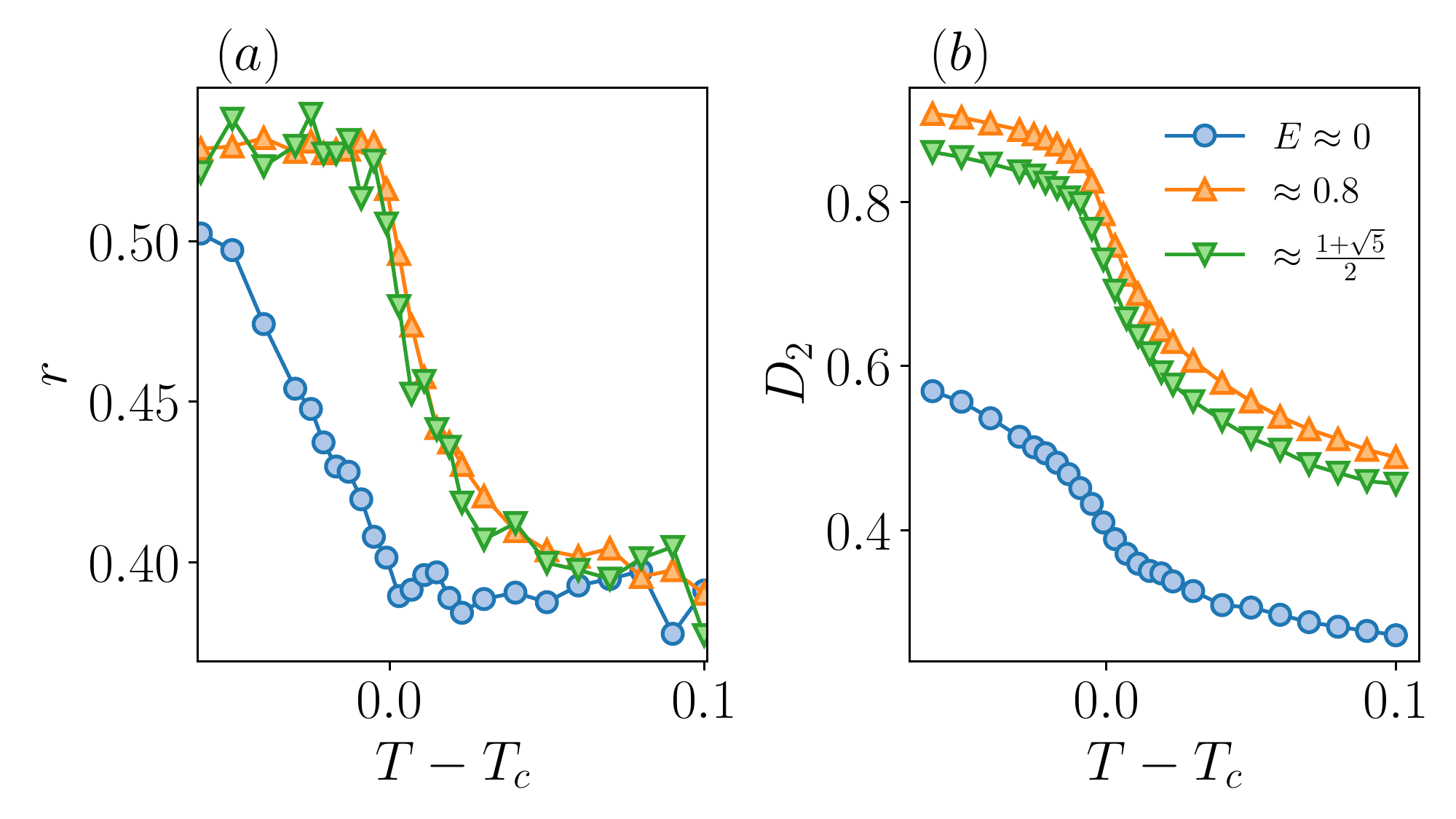}
    \caption{\textbf{Eigenvalue and eigenstate analysis for different energies (Ising interactions)}. (a): $r$-value statistics and (b): $D_2$ exponent for a fixed system size $L=1000$ and several energies $E$.}
    \label{fig:DiffE}
\end{figure}

Figures~\ref{fig:Pott3}~(a), (c), and (e) show the $r$-value, $D_2$, and $D_{1/2}$, respectively. As for the Ising correlated case (see main text), we observe a drastic change in behavior in localization properties when the critical temperature $T_c$ is traversed. In the ferromagnetic phase of the Potts model ($T<T_c$), the localization probes are consistent with the existence of extended states, while for $T>T_c$ the system appears localized. Once again, using the classical critical exponent $\nu \approx 0.46$ we obtain a good scaling collapse, as illustrated in Figs.~\ref{fig:Pott3}~(b),(d), and (f). This suggests the possible existence of a metal-insulator transition. As for the case with Ising interactions, it is important to point out that a different scenario remains possible, where the localization length becomes very large, but finite, once the classical phase transition is crossed. 

\section{Ising interaction}\label{app:ising}

In this section, we show further data for the Ising interacting case, Eq.~\eqref{eq:Hamiltonian}. In Fig.~\ref{fig:Eigen} we have shown $D_2$ and $D_{1/2}$ computed using Eq.~\eqref{eq:D_multi_exp} at $E\approx 0.8$. A different possibility to extract $D_2$ and $D_{1/2}$ is to fit directly $\langle \log{\IPR_q} \rangle \sim D_q(1-q)\log{L^2}$ to compute $D_q$. In Figs.~\ref{fig:Scaling1} (a) and (c) we show $\IPR_q$ as a function of system size for several temperatures. First, it is important to point out that for temperatures $T>T_c$, $\langle \log{\IPR_q} \rangle$ starts to bend as a function of $L$, as the system is localized. Fitting the last three points in Figs.~\ref{fig:Scaling1} (a) and (c), we can extract the multifractal exponents $D_2$ and $D_{1/2}$, respectively. In agreement with the results in Fig.~\ref{fig:Eigen} in the main text, $D_q \approx 1$ for $T<T_c$, consistent with the existence of an extended phase. 

In the main text, we focused our attention to energies $E\approx 0$ and $E\approx 0.8$. For the sake of completeness, in Fig.~\ref{fig:DiffE} we show the $r$-value and $D_2$ for a further energy $E\approx \frac{1+\sqrt{5}}{2}$. The results are similar to the ones for $E\approx 0.8$. 

\section{Mean-free path}\label{app:mean_free}

In this section, we estimate the enhancement of the mean free path, $\ell$, due to the presence of Ising correlations/interactions. This calculation focuses on the weak-scattering approximation, which assumes that scattering events due to the random potential are rare and weak; this in turn corresponds to $T\ll T_c$ and $W\ll 1$. The Hamiltonian $H = H_0 + V$ is decomposed into the free Hamiltonian $H_0$ and the disorder potential, $V = W \sum_{\vec{x}} \sigma_{\vec{x}} |\vec{x}\rangle \langle \vec{x} |$. 

The mean-free path is extracted from the self-energy, $\Sigma^{\pm}$, which is defined by the Dyson equation
\begin{equation}
    \langle G^{\pm} \rangle = G_0^{\pm} + G_0^{\pm} \Sigma^{\pm} \langle G^{\pm} \rangle
    \, , 
\end{equation}
where $\langle G^{\pm} \rangle = \langle (E^{\pm} - H)^{-1} \rangle$ and $G_0^{\pm} = (E^{\pm} - H_0)^{-1}$ are the disorder averaged Green functions for $H$ and $H_0$, respectively. $E^{\pm} = E\pm i\eta$, with $\eta>0$ an arbitrary infinitesimally small positive constant.  

The life-time $\tau$, which is  proportional to the mean-free path ($\tau \propto \ell$), is given by
\begin{equation}
    \frac{1}{\tau} = \pm 2 \Im (\Sigma^{\pm})
    \, .
\end{equation}
The self-energy can be approximated using the so-called first-order Born approximation (1BA)~\cite{Kramer1993,Bruus2003}, which is the lowest non-trivial order in perturbation theory: 
\begin{equation}
\Sigma^{\pm, 1BA} (\vec{k}) = \frac{W^2}{L^2}\sum_{\vec{k}'} C_{\text{Ising}}(\vec{k}-\vec{k}') G_0^{\pm}(\vec{k}')
\, ,
\end{equation}
where we have used the fact that $H_0$ is diagonal in the momentum basis; $L$ is the linear size of the system and $C_{\text{Ising}}$ is the averaged Fourier transform of the connected two-point correlation function $\langle \langle \sigma_{\vec{x}} \sigma_{\vec{0}}\rangle \rangle$. 

For $T<T_c$, the two-point correlation function can be approximated as $\langle \langle \sigma_{\vec{x}} \sigma_{\vec{0}}\rangle \rangle\sim M^2(T) e^{-x/\xi(T)}$~\footnote{Up to sub-leading polynomial corrections}, where $\xi^{-1}(T) = \log{\sinh{2/T}}$~\cite{McCoyWu_2013}. In this approximation, $C_{\text{Ising}}(k)$ is just a Lorentzian, $C_{\text{Ising}}(k) \propto \frac{M^2(T)\xi^{-1}}{(\xi^{-2} + k^2)}$.

The imaginary part of the free Green function $G_0^{\pm}$, in the weak scattering approximation, is proportional to the delta function, namely $\lim_{\eta \rightarrow 0} \Im [ G_0^\pm(E^{\pm}, \bf{k}') ] \propto \delta(k^2-k'^2)$, and 
\begin{equation}
\tau^{-1} \propto W^2 M^2(T)\int_0^{2\pi} d\theta \left[ A_1(\xi,\theta) \right]
\, ,
\end{equation}
with 
\begin{equation}
A_1(\xi,\theta) = \frac{\xi^{-1}}{2 k^2 ( 1-\cos\theta ) + \xi^{-2}}.
\end{equation}
Therefore,
\begin{equation}
\tau^{-1} \propto \frac{W^2 M^2(T)}{\sqrt{4 k^2 +\xi^{-2}}}.
\end{equation}
The result shows that $\tau\propto \ell$ scales as $(W^2 T)^{-1}$ in the limit $T\rightarrow 0$ and $W\rightarrow 0$, and diverges as expected.  

\bibliography{BIBLIO}

\end{document}